\documentclass[a4paper,11pt]{article}%
\usepackage{amsmath}
\usepackage{amsfonts}
\usepackage{amssymb}
\usepackage{graphicx}
\usepackage{pos}
\usepackage{mathrsfs}%
\providecommand{\U}[1]{\protect\rule{.1in}{.1in}}
\title{%
\vspace{-15mm}
\begin{minipage}{0.3\textwidth}
\normalsize 
KEK Preprint 2021-56 \\ 
CHIBA-EP-252 
\end{minipage}  \par \vspace{5mm}
Magnetic monopole dominance for the Wilson loops in higher representations%
}

\ShortTitle{Magnetic monopole dominance for the Wilson loops in higher representations}
\author*[a,b]{Akihiro Shibata}
\author[c]{Seikou Kato}
\author[d]{Kei-Ichi Kondo}
\affiliation[a]{ Computing Research Center, High Energy Accelerator Research Organization (KEK),  Tsukuba 305-0801, Japan}
\affiliation[b]{ Department of Accelerator Science, SOKENDAI (The Graduate Univercity for Advanced Studies), Tsukuba 305-0801, Japan}
\affiliation[c]{ Oyama National College of Technology, Oyama 323-0806, Japan}
\affiliation[d]{ Department of Physics, Graduate School of Science, Chiba University, Chiba 263-8522, Japan} 

\emailAdd{$^{a,b}$Akihiro.Shibata@kek.jp}
\emailAdd{$^{c}$skato@oyama-ct.ac.jp}
\emailAdd{$^{d}$kondok@faculty.chiba-u.jp}

\abstract{%
The dual superconductor picture is one of the most promising scenarios for quark confinement. 
To investigate this picture in a gauge-invariant manner, we have proposed a new formulation of 
Yang-Mills theory, named the decomposition method,  on the lattice. 
The so-called restricted field obtained from the gauge-covariant decomposition plays 
the dominant role in quark confinement. It has been known by preceding works 
that the restricted-field dominance is not observed for the Wilson loop in higher representations 
if the restricted part of the Wilson loop is obtained by adopting the Abelian projection 
or the field decomposition naively in the same way as done in the fundamental representation. 
Recently, through the non-Abelian Stokes theorem (NAST) for the Wilson loop operator, 
we have proposed suitable gauge-invariant operators constructed from the restricted field
 to reproduce the correct behavior of the original Wilson loop averages for higher representations.
 We have demonstrated the numerical evidence for the restricted-field dominance in the string tension,
 which means that the string tension extracted from the restricted part of the Wilson loop reproduces
 the string tension calculated from the original Wilson loop.
\par
 In this talk, we focus on the magnetic monopole. 
According to this picture, magnetic monopoles causing the dual superconductivity are the dominant degrees of freedom 
responsible for confinement. With the help of the NAST, we  define the magnetic monopole 
and the string tension extracted from the magnetic-monopole part of the Wilson loop in a gauge-invariant manner.
 We will further perform lattice simulations to measure the static potential for quarks 
in higher representations using the proposed operators and examine the magnetic monopole dominance 
in the string tension, which means that the string tension extracted from the magnetic-monopole part 
of the Wilson loop reproduces the proper string tension obtained from the original Wilson loop.
}
\FullConference{%
 The 38th International Symposium on Lattice Field Theory, LATTICE2021
  26th-30th July, 2021
  Zoom/Gather@Massachusetts Institute of Technology
}
\begin{document}
\maketitle

\section{Introduction}

The dual superconductor picture is one of the most promising scenarios for
quark confinement\cite{dualsuper}. According to this picture, magnetic
monopoles causing the dual superconductivity are regarded as the dominant
degrees of freedom responsible for confinement. However, it is not so easy to
establish this hypothesis. Indeed, even the definition of magnetic monopoles
in the pure Yang-Mills theory is not obvious. If magnetic charges are naively
defined from electric ones by exchanging the role of the magnetic field and
electric one according to the electric-magnetic duality, one needs to
introduce singularity to obtain non-vanishing magnetic charges, as represented
by the Dirac monopole. For such configuration, however, the energy becomes
divergent. Avoiding this issue in defining magnetic monopoles, there are two
prescriptions i.e., \emph{the Abelian projection method} and \emph{the
gauge-covariant decomposition method}.

The Abelian projection method, which is proposed by 't Hooft \cite{Hooft81},
is the most frequently used prescription. In the\emph{ }Abelian projection
method, the \textquotedblleft diagonal component\textquotedblright\ of the
Yang-Mills gauge field is identified with the Abelian gauge field, and a
magnetic monopole is defined as the Dirac monopole. The energy density of this
monopole can be finite everywhere because the contribution from the
singularity of a Dirac monopole can be canceled by that of the off-diagonal
components of the gauge field. For quarks in the fundamental representation,
indeed, such numerical simulations were already performed within the Abelian
projection using the maximal Abelian (MA) gauge in $SU(2)$\ and $SU(3)$%
\ Yang-Mills theories on the lattice\cite{SY90,SS94,STW02,SS14}. Then it was
confirmed that (i) the diagonal part extracted from the original gauge field
in the MA gauge reproduces the full string tension calculated from the
original Wilson loop average \cite{SY90,STW02,SS14}, which is called the
Abelian dominance, and that (ii) the monopole part extracted from the diagonal
part of the gauge field by applying the Toussaint-DeGrand procedure
\cite{DT80} mostly reproduces the full string tension \cite{SS94,STW02,SS14},
which is called the monopole dominance. However, it should be noted that the
MA gauge in the Abelian projection simultaneously breaks the local gauge
symmetry and the global color symmetry. This defect should be eliminated to
obtain the physical result by giving a procedure to guarantee the gauge invariance.

We have developed a new formulation of Yang-Mills theory, named the
decomposition method, which enables us to perform the numerical simulations on
the lattice in such a way that both the local gauge symmetry and the global
color symmetry remain intact, in sharp contrast to the Abelian projection
which breaks both symmetries (See \cite{KKSS15} for review). The
gauge-covariant decomposition was first proposed for $SU(2)$ by Cho
\cite{Cho80} and Duan and Ge \cite{DG79} independently, and later readdressed
by Faddeev and Niemi \cite{FN99a}, and developed by Shabanov \cite{Shabanov99}
and the Chiba University group \cite{KMS05,KMS06,Kondo06}. In this method, the
gauge field is decomposed into two parts; A part called restricted field
transforms under the gauge transformation just like the original gauge field,
while the other part called the remaining field transforms like an adjoint
matter. The key ingredient in this decomposition is the Lie-algebra valued
field with a unit length that we call the color field. The decomposition is
constructed in such a way that the field strength of the restricted field is
\textquotedblleft parallel\textquotedblright to the color field. Then
monopoles can be defined by using the gauge-invariant part proportional to the
color field in the field strength just as the Abelian field strength in the
Abelian projection. While the main advantage of the field decomposition is its
gauge covariance, another advantage is that, through a version of the
non-Abelian Stokes theorem (NAST) invented originally by Diakonov and
Petrov\cite{DP89} and extended in a unified way in
\cite{KondoIV,KT00b,KT00,Kondo08,Kondo08b}, the restricted field naturally
appears in the surface-integral representation of the Wilson loop. By virtue
of this method, we understand how monopoles contribute to the Wilson loop at
least classically. The field decomposition was extended to $SU(N)$ $(N\geq3)$
gauge field in \cite{KMS05,KMS06,BCK02,KSM08,BCK02}. By way of the non-Abelian
Stokes theorem for the Wilson loop operator \cite{Kondo08,Kondo08b}, indeed,
it was found that the different type of decomposition called the minimal
option is available for $SU(3)$ and $SU(N)$ for $N\geq4$ \cite{KMS06,KSM08}.

For quarks in the fundamental representation, we have presented the lattice
formulation and demonstrated some numerical evidences for the dual
superconductivity by using the gauge-covariant decomposition method
\cite{IKKMSS06,KSSMKI08,SKS10,KSSK11,SKKS16,KKS15,CC}. This framework improves
the Abelian projection in the gauge-independent manner. Moreover, the MA gauge
in the Abelian projection is not the only way to recover the string tension in
the fundamental representation. Even for the minimal option, we have
demonstrated the restricted field dominance and monopole dominance in the
string tension for quarks in the fundamental representation
\cite{KSSK11,SKKS16,KKS15}. Thus, our method enables one to extract various
degrees of freedom to be responsible for quark confinement by combining the
option of gauge-covariant field decomposition and the choice of the reduction
condition, which is not restricted to the Abelian projection and the MA gauge, respectively.

As for quarks in higher representations, it was known by preceding works that
the restricted field dominance is not observed for the Wilson loop in higher
representations if the restricted part of the Wilson loop is extracted by
adopting the Abelian projection or the field decomposition naively in the same
way as in the fundamental representation \cite{DFGO96}. This is because, in
higher representations, the diagonal part of the Wilson loop does not behave
in the same way as the original Wilson loop. Poulis heuristically found the
correct way to extend the Abelian projection approach for the adjoint
representation in the $SU(2)$ Yang-Mills theory \cite{Poulis96}. Recently, the
NAST for the Wilson loop operator are extended to any representations
\cite{MK15}. By virtue the NAST, we have proposed suitable gauge-invariant
operators constructed from the restricted field to reproduce the correct
behavior of the original Wilson loop averages for higher representations. We
have further demonstrated the numerical evidence that the proposed operators
well reproduce the behavior of the original Wilson loop average, namely, the
linear part of the static potential with the correct value of the string
tension \cite{MSKK2019}.

In this talk, we focus on the magnetic monopole. With the help of the NAST,~we
can define the magnetic monopole and the string tension extracted from the
magnetic-monopole part of the Wilson loop in a gauge-invariant manner. We will
further perform lattice simulations to measure the static potential for quarks
in higher representations using the proposed operators and examine the
magnetic monopole dominance in the string tension.

\section{The gauge-covariant field decomposition}

We decompose the gauge link variable $U_{x,\mu}$ into the product of the two
variables $V_{x,\mu}$ and $X_{x,\mu}$ in such a way that the new variable
$V_{x,\mu}$ is transformed by the full $SU(N)$ gauge transformation
$\Omega_{x}$ as the gauge link variable $U_{x,\mu}$, while $X_{x,\mu}$
transforms as the site variable:
\begin{subequations}
\begin{align}
&  U_{x,\mu}=X_{x,\mu}V_{x,\mu}\in G=SU(N),\qquad U_{x,\mu}\longrightarrow
U_{x,\nu}^{\prime}=\Omega_{x}U_{x,\mu}\Omega_{x+\mu}^{\dag},\text{ }\\
&  V_{x,\mu}\longrightarrow V_{x,\nu}^{\prime}=\Omega_{x}V_{x,\mu}%
\Omega_{x+\mu}^{\dag},\qquad X_{x,\mu}\longrightarrow X_{x,\nu}^{\prime
}=\Omega_{x}X_{x,\mu}\Omega_{x}^{\dag}.
\end{align}
From the physical point of view, $V_{x,\mu}$ which we call the restricted
field, could be the dominant mode for quark confinement, while $X_{x,\mu}$ is
the remainder part. The possible options of the decomposition are
discriminated by the stability subgroup of the gauge group. Here, we only
consider the maximal option.

The maximal option is obtained for the stability subgroup of the maximal torus
subgroup of $G$: $\tilde{H}=U(1)^{N-1}\subset SU(N).$ The resulting
decomposition is the gauge-invariant extension of the Abelian projection in
the MA gauge. \ We introduce color fields as,
\end{subequations}
\begin{equation}
\mathbf{n}^{(k)}(x)=\Theta(x)H_{k}\Theta^{\dag}(x)\in Lie[G/\tilde{H}]\text{
\ \ \ \ \ \ }(k=1,\ldots,N-1)\text{,}%
\end{equation}
which are expressed using a common $SU(N)$-valued field $\Theta(x)$ with the
Cartan generators $H_{k}$. The decomposition is obtained by solving the
defining equations:
\begin{subequations}
\begin{align}
D_{\mu}^{\epsilon}[V]\mathbf{n}_{x}^{(k)}  &  :=\frac{1}{\epsilon}\left[
V_{x,\mu}\mathbf{n}_{x+\mu}^{(k)}-\mathbf{n}_{x}^{(k)}V_{x,\mu}\right]
=0\text{ },\label{eq:define-max}\\
\text{\ }g_{x}  &  :=e^{i2\pi q/N}\exp\left(  -i\sum_{j=1}^{N-1}a_{x}%
^{(j)}\mathbf{n}_{x}^{(j)}\right)  \text{,} \label{eq:define-resdidual}%
\end{align}
where, the variable $g_{x}$ is the $U(1)^{N-1}$ part which is undetermined
from Eq.(\ref{eq:define-max}) alone, $a_{x}^{(j)}$ are coefficients, and $q$
is an integer. Note that the above defining equations correspond to the
continuum version ($g_{x}=\mathbf{1}$): $D_{\mu}[%
%TCIMACRO{\TeXButton{mathscr}{\mathscr}}%
%BeginExpansion
\mathscr
%EndExpansion
V]\mathbf{n}^{(k)}(x)=0$ and $\mathrm{tr}(%
%TCIMACRO{\TeXButton{mathsrc}{\mathscr}}%
%BeginExpansion
\mathscr
%EndExpansion
X_{\mu}(x)\mathbf{n}^{(k)}(x))$ $=0$, respectively. These defining equations
can be solved exactly \cite{SKS10}, and the solution is given by
\end{subequations}
\begin{align}
X_{x,\mu}  &  =\widehat{K}_{x,\mu}^{\dag}\det(\widehat{K}_{x,\mu})^{1/N}%
g_{x}^{-1},\text{ \ \ \ \ \ \ \ \ \ }V_{x,\mu}=X_{x,\mu}^{\dag}U_{x,\mu
},\nonumber\\
\widehat{K}_{x,\mu}  &  :=\left(  K_{x,\mu}K_{x,\mu}^{\dag}\right)
^{-1/2}K_{x,\mu},\text{ \ \ \ \ \ \ \ \ \ \ \ }K_{x,\mu}:={1}+2N\sum
_{k=1}^{N-1}\mathbf{n}_{x}^{(k)}U_{x,\mu}\mathbf{n}_{x+\mu}^{(k)}U_{x,\mu
}^{\dag}. \label{eq:decomp-max}%
\end{align}
In the naive continuum limit, we can reproduce the decomposition in the
continuum theory.

\section{Non-Abelian Stokes theorem and magnetic monopole}

It is known that if we adopt the Abelian projection naively to higher
representations, the monopole contributions do not reproduce the correct
behavior \cite{DFGO96}. Thus, we have to find a more appropriate way to
extract the monopole contributions. We can relate the decomposed field
variables to a Wilson loop operator through a version of the NAST which was
proposed by Diakonov and Petrov \cite{DP89}. Recently, \ the NAST\ is extended
to the Wilson loop operator in any representations for $SU(N)$ Yang-Mills
theory \cite{MK15}. Through the NAST, therefore, we have proposed suitable
gauge-invariant operators constructed from the restricted field to reproduce
the correct behavior of the original Wilson loop averages for higher
representations, and demonstrated the restricted field dominance in the string
tension \cite{SMKK2018,MSKK2019}. Here, we focus on the magnetic monopole
contribution. We quickly summarize the result of the Ref. \cite{MK15},

The Wilson loop operator in a representation $R$ of $SU(N)$ Yang-Mills theory
is rewritten into the surface integral form by introducing a functional
integral on the surface $\Sigma$ surrounded by the loop $C$ as
\begin{align}
W_{C}[%
%TCIMACRO{\TeXButton{mathscr{A}}{\mathscr{A}}}%
%BeginExpansion
\mathscr{A}%
%EndExpansion
]=  &  \int[d\mu(g)]_{\Sigma}\exp\left[  -ig_{{}_{\mathrm{YM}}}\sqrt
{\frac{N-1}{2N}}\int_{\Sigma:\partial\Sigma=C}f^{g}\right]  ,\quad
f^{g}:=\frac{1}{2}f_{\mu\nu}^{g}(x)dx^{\mu}\wedge dx^{\nu},\nonumber\\
f_{\mu\nu}^{g}(x)=  &  \kappa(\partial_{\mu}\mathrm{tr}%
\{\boldmath n(x)\mathscr{A}_{\nu}(x)\}-\partial_{\nu}\mathrm{tr}%
\{\boldmath n(x)\mathscr{A}_{\mu}(x)\}+ig_{{}_{\mathrm{YM}}}^{-1}%
\mathrm{tr}\{\boldmath n(x)[\partial_{\mu}\boldmath n_{k}(x),\partial_{\nu
}\boldmath n_{k}(x)]\}),\nonumber\\
=  &  \partial_{\mu}\{{n}^{A}(x)\mathscr{A}_{\nu}^{A}(x)\}-\partial_{\nu}%
\{{n}^{A}(x)\mathscr{A}_{\mu}^{A}(x)\}-g_{{}_{\mathrm{YM}}}^{-1}f^{ABC}{n}%
^{A}(x)\partial_{\mu}{n}_{k}^{B}(x)\partial_{\nu}{n}_{k}^{C}(x),\nonumber\\
\boldmath n(x)=  &  \sqrt{\frac{2N}{N-1}}\Lambda_{j}\boldmath n_{j}%
(x),\quad\boldmath n_{j}(x)=g(x)H_{j}g^{\dagger}(x)\quad(j=1,...,r).
\label{eq:NAST}%
\end{align}
where $[d\mu(g)]_{\Sigma}$ is the product of the Haar measure over the surface
$\Sigma$, $\Lambda_{k}$ is the $k$-th component of the highest-weight of the
representation $R$. It should be noticed that the Wilson loop operator is
represented by using the restricted field $\mathscr V$. We choose the
highest-weight state as the reference state. For $G=SU(2)$ the highest-weight
vector of the representation with the spin $J$ is given by $\vec{\Lambda
}_{SU(2)}=(\Lambda_{3})=(J)=(\frac{m}{2})$, while for $G=SU(3)$ the
highest-weight vector of the representation with the Dynkin indices $[m,n]$ is
given by $\vec{\Lambda}=(\Lambda_{3},\Lambda_{8})=\left(  \frac{m}{2}%
,\frac{m+2n}{2\sqrt{3}}\right)  .$

Through the NAST we can define the gauge-invariant magnetic-monopole from the
field strength in the same manner as the Dirac monopoles for the Abelian-like
gauge-invariant field. \ By using the Hodge decomposition for the field
strength, the Wilson loop operator in arbitrary representation is written in
terms of the electric current $j$ and the magnetic current $k$:
\begin{equation}
W_{C}[\mathscr{A}]=\int[d\mu({g})]\exp\left\{  -ig_{{}_{\mathrm{YM}}}%
\sqrt{\frac{N-1}{2N}}[(\omega_{\Sigma_{C}},k)+(N_{\Sigma_{C}},j)]\right\}  ,
\label{eq:NAST-Hdcomp}%
\end{equation}
where we have defined the $(D-3)$-form $k$ and the one-form $j$ in $D$
space-time dimensions:
\begin{equation}
k:=\delta{}^{\displaystyle\ast}f^{g},\quad j:=\delta f^{g},
\end{equation}
we have introduced an antisymmetric tensor $\Theta_{\Sigma_{C}}$ of rank two
which has the support only on the surface $\Sigma_{C}$ spanned by the loop
$C$:
\[
\Theta_{\Sigma_{C}}^{\mu\nu}(x):=\int_{\Sigma_{C}:\partial\Sigma_{C}=C}%
d^{2}S^{\mu\nu}(x(\sigma))\delta^{D}(x-x(\sigma)),
\]
and we have defined the $(D-3)$-form $\omega_{\Sigma_{C}}$ and one-form
$N_{\Sigma_{C}}$ using the Laplacian $\Delta$ by
\begin{equation}
\omega_{\Sigma_{C}}:={}^{\displaystyle\ast}d\Delta^{-1}\Theta_{\Sigma_{C}%
}=\delta\Delta^{-1}{}^{\displaystyle\ast}\Theta_{\Sigma_{C}},\quad
N_{\Sigma_{C}}:=\delta\Delta^{-1}\Theta_{\Sigma_{C}},
\end{equation}
with the inner product for two forms being defined by
\begin{align}
&  (\omega_{\Sigma_{C}},k)=\frac{1}{(D-3)!}\int d^{D}xk^{\mu_{1}\cdots
\mu_{D-3}}(x)\omega_{\Sigma_{C}}^{\mu_{1}\cdots\mu_{D-3}}(x),\nonumber\\
&  (N_{\Sigma_{C}},j)=\int d^{D}xj^{\mu}(x)N_{\Sigma_{C}}^{\mu}(x).
\end{align}

Therefore we can construct the proper Wilson-loop operator by using the
decomposed variable $V$ in Eq(\ref{eq:decomp-max}) on the lattice.
Incidentally, we can show \cite{KKSS15} that the gauge-invariant field
strength $F_{\mu\nu}^{g}$ is equal to the component of the non-Abelian field
strength $\mathscr{F}[\mathscr{V}]$ of the restricted field $\mathscr{V}$ (in
the decomposition $\mathscr{A}=\mathscr{V}+\mathscr{X}$) projected to the
color field ${\boldmath n}$:
\begin{align}
&  W_{R}[\mathscr A;C]:=\int[d\mu(g)]_{\Sigma}\exp\left(  ig\int
_{\Sigma:\partial\Sigma=C}dS_{\mu\nu}\sum_{k=1}^{N-1}\Lambda_{k}F_{\mu\nu
}^{(k)}\right)  \text{ ,}\nonumber\\
&  F_{\mu\nu}^{g}=\mathrm{tr}\{{\boldmath m}\mathscr{F}_{\mu\nu}%
[\mathscr{V}]\}=\Lambda_{j}f_{\mu\nu}^{(j)},\quad f_{\mu\nu}^{(j)}%
=\mathrm{tr}\{{\boldmath n}_{j}%
%TCIMACRO{\TeXButton{mathscr F[v]}{\mathscr{F} [\mathscr{V}]}}%
%BeginExpansion
\mathscr{F} [\mathscr{V}]%
%EndExpansion
\}\text{ .} \label{eq;NAST2}%
\end{align}
This relation is useful in calculating the contribution from magnetic
monopoles to the Wilson loop average from the viewpoint of the dual
superconductor picture for quark confinement. We might identify the restricted
field $\mathscr{V}$ and the color field ${\boldmath n}_{j}$ in
Eq(\ref{eq;NAST2}) with the restricted field and the color field in
Eq(\ref{eq:decomp-max}), respectively. However, since the NAST
Eq(\ref{eq;NAST2}) includes the integration over the measure $[d\mu
(g)]_{\Sigma}$, which indicates the integral over the whole directions of
color fields, the field correspondences are not simple. In the gauge-covariant
decomposition method, it should be remined that the color fields are
introduced as\ auxiliary fields and the decomposition is defined in the space
with extended symmetry $G\times G/\tilde{H}$. In order to make the theory
written by the decomposed fields equivalent to the original Yang-Mills theory
with the same gauge symmetry, it is necessary to impose a reduction condition
between the color field and the Yang-Mills field. Thus, this requirement can
be formulated as a path integral with the reduction condition imposed as a
constraint. While in the lattice measurements this means that the operator is
evaluated by using the color field which is obtained as the solution of the
reduction condition in place of performing an integral over the color field.

Thus we define the magnetic monopole by using the restricted field:%
\begin{align}
&  k_{\mu}=\sum_{j=1}^{r}\Lambda_{j}k_{\mu}^{(j)}\text{ ,\ \ \ \ \ \ }k_{\mu
}^{(j)}=\frac{1}{2}\epsilon^{\mu\nu\rho\sigma}\partial_{\nu}%
%TCIMACRO{\TeXButton{mathscr F}{\mathscr{F}}}%
%BeginExpansion
\mathscr{F}%
%EndExpansion
_{\alpha\beta}^{(j)}[V]\text{ ,}\nonumber\\
&  V_{x,\mu}V_{x+\hat{\mu},\nu}V_{x,+\hat{\nu},\mu}^{\dag}V_{x,\nu}^{\dag
}=\exp\left(  -i%
%TCIMACRO{\TeXButton{mathscr F}{\mathscr{F}}}%
%BeginExpansion
\mathscr{F}%
%EndExpansion
_{\alpha\beta}[V]\right)  =\exp\left(  -i\sum_{k=1}^{r}%
%TCIMACRO{\TeXButton{mathscr F}{\mathscr{F}}}%
%BeginExpansion
\mathscr{F}%
%EndExpansion
_{\alpha\beta}^{(k)}[V]\mathbf{n}_{x}^{(k)}\right)  \text{ .}%
\end{align}

\section{Lattice data}

It should be examined on the lattice whether or not these monopoles can
reproduce the expected infrared behavior of the original Wilson loop average
in the higher representation.\ As the first step, we examine the case of the
$SU(2)$ Yang-Mills theory. We set up the gauge configuration for the standard
Wilson action at $\beta=2.5$ on the $24^{4}$ lattice for $SU(2)$ case. We
prepare $500$ configurations every $100$ sweeps after $3000$ thermalization by
using the pseudo heat-bath method. The restricted field is obtained by using
the decomposition formula Eq(\ref{eq:decomp-max}) for a given set of
configurations of the Yang-Mills field $\{U_{x,\mu}\}$ and the color field
$\{\mathbf{n}_{x};\mathbf{n}_{x}:=\Theta_{x}T^{3}\Theta_{x}^{\dag}$ $\}$,
where the color field is determined by minimizing the reduction condition
functional:
\begin{equation}
R[U,\{{n}\}]=\sum_{x,\mu}\operatorname{tr}[(D_{\mu}[U]\mathbf{n}_{x})^{\dag
}(D_{\mu}[U]\mathbf{n}_{x})]\quad. \label{eq:red_con}%
\end{equation}

We calculate the Wilson loop average $W(R,T)$ for the rectangular loop with
length $T$ and width $R$ to derive the potential $V(R,T)$ through the formula
$V(R,T):=-\log(W(R,T+1)/W(R,T)$. In the measurement of the Wilson loop average
we apply the APE smearing to reduce noises and eliminate exciting modes. The
Wilson-loop operator for the spin-$J$ representation is given by
\begin{subequations}
\begin{align}
W_{[J]}^{SU(2)}[U;C] &  =\sum_{k=0}^{\left\lfloor J\right\rfloor
}\operatorname{tr}(U_{C}^{2(j-k)})\text{\qquad with }U_{C}:=\prod
_{\left\langle x,\mu\right\rangle \in C}U_{x,\mu}\text{ , }\label{eq:WloopYM}%
\\
W_{[J]}^{SU(2)}[V;C] &  =\frac{1}{2}\operatorname{tr}(V_{C}^{2J})\text{\qquad
with }V_{C}:=\prod_{\left\langle x,\mu\right\rangle \in C}V_{x,\mu}\text{
.}\label{eq:WloopV}%
\end{align}
For $SU(2)$ case the magnetic monopole part of Eq(\ref{eq:NAST-Hdcomp}) can be
calculated as
\end{subequations}
\begin{align}
&  W_{[J]}^{SU(2)}[k;C]=\exp\left\{  2\pi iJ\sum_{x,\mu}k_{\mu}(x)N_{\mu
}(x)\right\}  ,\nonumber\\
&  k_{\mu}(x)=\frac{1}{2}\epsilon^{\mu\nu\alpha\beta}\partial_{\nu}%
\Theta_{\alpha\beta}(x)\text{, \ \ \ \ \ }\Theta_{\alpha\beta}=\arg
\text{tr}\left(  (\mathbf{1}+2\mathbf{n}_{x})V_{x,\mu}V_{x+\hat{\mu},\nu
}V_{x,+\hat{\nu},\mu}^{\dag}V_{x,\nu}^{\dag}\right)  \text{,}\nonumber\\
&  N_{\mu}(x)=\sum_{y}\Delta^{-1}(x-y)\frac{1}{2}\epsilon_{\mu\alpha
\beta\gamma}\partial_{\alpha}S_{\beta\gamma}^{C}(y)\text{,}\label{eq:NAST-Mon}%
\end{align}
where $\Delta^{-1}(x-y)$ is the inverse of the Laplacian, and $S_{\beta\gamma
}^{C}(y)$ is the surface element with the Wilson loop ($C$) as boundary which
satisfies the equation $\partial_{\beta}S_{\beta\gamma}^{C}(y)=C_{\gamma}(y)$
with $C_{\gamma}(y)$ being the tangent vector of the path $C.$%

%TCIMACRO{\TeXButton{B}{\begin{figure}[tbp] \centering}}%
%BeginExpansion
\begin{figure}[tbp] \centering
%EndExpansion%
%TCIMACRO{\TeXButton{potential_nlcv_adjm_b250l24t6r12_fit.eps}{\includegraphics
%[width=10cm]{potential_nlcv_adjm_b250l24t6r12_fit.eps}}}%
%BeginExpansion
\includegraphics[width=10cm]{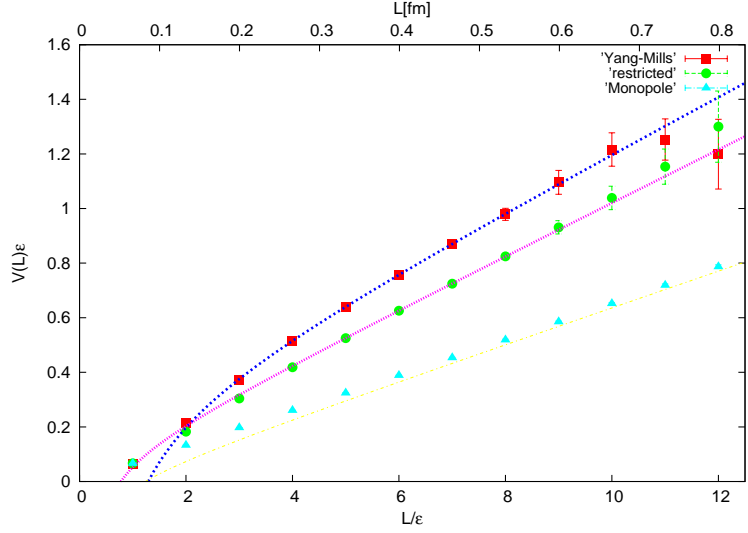}%
%EndExpansion
\caption{The static potential for quarks in the adjoint representation.
}\label{fig:potential}%
%TCIMACRO{\TeXButton{E}{\end{figure}}}%
%BeginExpansion
\end{figure}%
%EndExpansion

Figure \ref{fig:potential} shows the preliminary result of the statical
potential for a pair of quark and antiquark in the adjoint representation
($J=1)$, where we simultaneously plot the the statical potential calculated
from the magnetic monopole Eq(\ref{eq:NAST-Mon}), the original Yang-Mills
field Eq(\ref{eq:WloopYM}), and the restricted field Eq(\ref{eq:WloopV}). We
find the magnetic monopole play a dominant role in the string tension
($\sigma_{mono}/\sigma_{full}\simeq68\%$) for the quarks in the adjoint
representation as well as the restricted field dominance in the string tension
which we have already shown in Ref.\cite{SMKK2018,MSKK2019}.

\section{Summary and Discussion}

We have investigated the magnetic monopole contribution for quark confinement
in the higher representation. Through the non-Abelin Stokes theorem for the
Wilson loop in the higher representation, we have defined the magnetic
monopole and the string tension extracted from the magnetic-monopole part of
the Wilson loop in the higher representation in a gauge-invariant manner. We
have performed lattice simulations for the $SU(2)$ Yang-Mills theory and
measured the Wilson loop average in the adjoint representation calculated from
the magnetic monopole to examine the magnetic-monopole dominance in the string
tension. We have found that the magnetic monopole plays a dominant role in the
string tension for the quarks in the adjoint representation.

To examine the restricted-field dominance and the magnetic-monopole dominance
in the string tension for the quarks in the higher representations, we need to
investigate the Wilson loops in various gauge groups and various
representations. We will further investigate large size Wilson loops to
examine whether or not the string breaking for Wilson loops in the higher
representation calculated from the original Yang-Mills field can be reproduced
for the Wilson loops calculated from the restricted field as well from the
magnetic monopole.

.

\section*{Acknowledgement}

This work was supported by Grant-in-Aid for Scientific Research, JSPS KAKENHI
Grant Number (C) No.19K03840. The numerical calculations were performed \ in
part using Cygnus and Oakforest-PACS at the CCS, University of Tsukuba.

\end{document}